\documentclass[11pt,a4paper]{article}
\usepackage[english]{babel}
\usepackage{algorithm}
\usepackage{subfig}
\usepackage{algpseudocode}
\usepackage{authblk}

\usepackage[a4paper,top=3cm,bottom=2cm,left=3cm,right=3cm,marginparwidth=1.75cm]{geometry}

\usepackage{amsmath}
\usepackage{graphicx}
\usepackage{braket}

\title{Continuous variable quantum key distribution with multi-mode signals for noisy detectors}
\author[1]{Rupesh Kumar\thanks{Corresponding Author: rupesh.kumar@york.ac.uk}}
\author[2]{Xinke Tang}
\author[2]{Adrian Wonfor}
\author[2]{Richard Penty }
\author[2]{Ian White }
\affil[1]{Quantum Communications Hub,  University of York,York YO10 5DD, UK}
\affil[2]{ Centre for Photonic Systems, University of Cambridge, Cambridge CB3 0FA, UK}

\begin{document}
\date{}
\maketitle
\begin{abstract}
This paper proposes a multi-mode Gaussian modulated continuous variable quantum key distribution (CV-QKD) scheme able to operate at high bandwidth despite using conventional noisy, coherent detectors. We demonstrate enhancement in shot-noise sensitivity as well as reduction in the electronic noise variance of the coherent receiver of the multi-mode CV-QKD system. A proof-of-concept simulation is presented using  multiple modes; this demonstrates an increase in signal-to-noise ratio and secure key rate at various transmission  distances with increasing signal modes 
\end{abstract}
\section{Introduction}
Quantum Key Distribution (QKD) \cite{Bennett1984} is a promising technology for sharing information with unconditional security.  Continuous-variable QKD (CV-QKD) \cite{Grosshans2002,Grosshans2003,Weedbrook2004,Diamanti2015} based on quadrature modulation of light is attractive, as it offers higher efficiency in secure key-generation rates (especially when used in dense wavelength division multiplexing networks \cite{Kumar2015}) compared with discrete variable QKD (DV-QKD), which is based  on single-photon encoding and detection \cite{Scarani2009,Gisin2002}. CV-QKD is  also very suitable for photonic integration, as the transceivers  are very similar to those in classical coherent high-speed  communication systems \cite{Orieux2016,Kikuchi2016}. However, classical coherent  detectors at higher bandwidth exhibit higher electronic noise  compared with the fundamental vacuum noise—shot noise  \cite{Zhang2012}. This limits the shot-noise sensitivity of the detector,  which is a fundamental requirement for detecting quadrature  modulated quantum signals. Higher electronic noise limits the  performance of CV-QKD systems as it can reduce the signal to-noise ratio (SNR), especially at longer transmission distances  \cite{Jouguet2013a}. It is possible to use an intense optical local oscillator (LO)  to cause the shot noise to dominate the Johnson thermal electronic noise—the main source of electronic noise in the coherent detector \cite{Zhang2012}. However, this is not achievable at higher  repetition rates due to laser power restrictions in transmitted  LO (TLO) CV-QKD systems. An alternative approach, generating LO locally (LLO), requires a pair of low-linewidth  laser sources whose phase-mismatching error with respect to  each other creates additional noise that limits long-distance  performance \cite{Qi2015,Soh2015,Adrien2017}. 

Here, we propose a new scheme for CV-QKD with a multi- mode Gaussian modulated protocol  \cite{Grosshans2002,Weedbrook2004} that exhibits higher  SNR compared to a conventional high-bandwidth CV-QKD  system. Multi-mode signals in CV-QKD have been considered  in \cite{ Fang2014, Usenko2014, Derkach2017} for Gaussian as well as in  \cite{Qu2017a,Qu2017b} for discrete modulation protocols. The basic feature of our scheme involves the  transmission of multi-mode, mutually incoherent signals prepared by Alice and their joint measurement at once using a  coherent detector by Bob. Here, by the term “mode,” we mean orthogonal states in any degrees of freedom of the light. We  have selected the wavelength, as it is practically easier to generate different modes and is not as limited in number of degrees  compared to other modes such as polarization.

This paper is organized as follows. In Section 2, we describe  the conventional Gaussian modulation CVQKD protocol and  a set of equations useful for the estimation of final secure key  rate. In Section 3, we consider the basic assumptions of the  detector for performing CV-QKD. A multi-mode scheme  for enhancing the SNR will be discussed in Section 4, and principle demonstration and simulation results as to the gain in secure key rate will be given in Section 5. Conclusions will be  given in Section 6.

\section{Gaussian modulated  CV-QKD}

In a Gaussian modulated CV-QKD system, Alice prepares  coherent state $\ket{\alpha_A} = \ket{X_A + i P_A}$ and sends to Bob through a quantum channel. The quadratures, $X_A$ and $P_A$, are drawn  from two sets of normally distributed random variables $\mathcal{N}\{0,V_A\}$, of zero mean and variance $V_A$. Bob measures the quadratures  with respect to a reference signal- LO,  using a shot noise limited homodyne \cite{Grosshans2002} or heterodyne  \cite{Weedbrook2004} coherent receiver. Without loss of generality, we consider a homodyne detection scheme where Bob randomly measures one of the quadratures, $X_B$ or $P_B$. The LO can be either transmitted from Alice (TLO)\cite{Lodewyck2007} or generate locally at Bob (LLO) from a second laser \cite{Qi2015,Soh2015}.  In the latter scheme, Alice also needs to send a reference signal to lock the phase of the laser at Bob. The quantum channel is characterized  by two parameters, transmittance $T$ and excess noise $\xi$, which can be estimated using the following equations:

\begin{eqnarray}
\label{Cov}
<X_A X_B>   &= & \sqrt{\eta T} V_A\\\label{Var}
var(X_B)= V_B &=& \eta T V_A + N_0+ \eta T \xi+ v_{ele}
\end{eqnarray}

Where, $\eta$ is the detection efficiency of Bob, $N_0$ is the shot noise variance and $v_{ele}$ is the electronic noise variance expressed in shot noise unit (snu or otherwise labeled as $N_0$). The l.h.s of  Eq.\eqref{Cov} is the covariance between $X_A$ and $X_B$. Above equations hold true for $P_B$ quadrature measurements as well.  With the help of data reconciliation, in particular reverse reconciliation\cite{Grosshans2003a}  for channels with loss greater than 3dB, Alice and Bob can extract secure keys from correlated quadrature values $\{X^1_A..X^{N/3}_A, X^1_B..X^{N/3}_B\}$ by preforming error correction and privacy amplification\cite{Jouguet2011}. Here we assume one third of the  total number of $N$ measured quadratures values are used for channel parameter estimation and one third is for real-time shot noise variance measurement. 
From the estimated values of  the channel parameters, $T$ and $\xi$, secure key generation rate can be estimated. Secure key rate  in asymptotic  limit, under collective attack, can be written as \cite{Grosshans2005}:

\begin{equation}
K= \gamma (\beta I_{AB} - \min \{\chi_{EB},\chi_{EA} \})
\label{Keyrate}
\end{equation}

Here, $\gamma$ is the fraction of quadrature data used for secure key generation and $\beta$ is the reconciliation efficiency. $I_{AB}$ is the  mutual information between Alice and Bob and $ \chi_{EA},\chi_{EB} $  are the Holevo bounds to Eve’s accessible information \cite{Weedbrook2012a}, for the quadrature  prepared by Alice and  Bob's quadrature measurement outcomes, respectively. $ \chi_{EA}$ pertains to the direct reconciliation where  Bob corrects his noisy measurement outcomes with respect to Alice.  While in reverse reconciliation, Alice corrects initial quadrature information as per Bob's  noisy measurement outcomes. In this case, Eve is forced to gather knowledge about Bob's measurements, $\chi_{EB}$, in order to maximize her eavesdropping.
In reverse reconciliation, the noise in Bob's measurements improves the robustness of the protocol to the channel excess noise by partly decoupling eavesdropper  from the Bob's measurement outcomes and helps to extend the transmission distance beyond 3dB limit in the case of direct reconciliation. The total noise,  $\chi_{tot} =\chi_{line} + \chi_{hom}/T$, in the CV-QKD system is separated into two. The channel noise $\chi_{line} = (1-T)/T +\xi$ and detection noise $\chi_{hom} = (1-\eta + v_{ele})/\eta$. It is assumed that the  electronic noise variance, $v_{ele}$, and detection efficiency, $\eta$,  cannot be accesed by Eve so that both can be calibrated and  trusted \cite{Lodewyck2007}. This applies  in the  reverse reconciliation  procedure and so we restrict our calculation to it. Here, $v_{ele}$ plays a role in SNR.  One can find the signal to noise ratio in the system as: $SNR = V_A/ (1+\chi_{tot})$ and then estimate the mutual information $I_{AB}$ from the following equation:

\begin{equation}
 I_{AB}  = \frac{1}{2} log_2(1+SNR)
 \label{Iab}
\end{equation}

In order to estimate the Holevo bound for Eve's accessible  information,  $\chi_{EB}$, under collective attack on the Gaussian states sent by Alice, one can use the equation below:

\begin{equation}
 \chi_{EB}  =  S(\rho_E) - \int p(X_B)S(\rho_{E|B})dX_B
  \label{CHIBE}  
\end{equation}

where,  $S(\rho_E)$ is the Von Neumann entropy of the state that Eve poses, $p(X_B)$ is the probability distribution of Bob's measurements, and $\rho_{E|B}$ is Eve's states conditioned on Bob's measurement. Please refer to appendix A for further description on Eq.\eqref{CHIBE}. From the parameters estimation, if the level of excess noise  $\xi$  is below the null key threshold, Alice and Bob can estimate the secure key rate, Eq.\eqref{Keyrate}, and proceed to error correction and privacy amplification in order to generate unconditional secure keys.

\section{Prerequisite for  performing Gaussian modulated CV-QKD }

In order to guarantee the security of a protocol, it is necessary to put forward assumptions on the underlying hardware systems and the way they perform. In CV-QKD protocols,  quantum uncertainty  imposed by the shot noise on the quadrature measurements  provides the fundamentals to the theoretical security    \cite{Grosshans2002}. For this, it is assumed that the coherent receiver that detects the quadratures does have adequate sensitivity to  infer the quantum uncertainty from the measurement outcomes. Such receivers are referred as "shot  noise  limited". 
It is also assumed that the transfer function of the quantum channel, together with subsequent detection by the  detector, follows a linear Gaussian model such that the variance of the output signal as well as the induced noises are Gaussian and linear with respect to the input signal. This requires a receiver with linear response function where the variance of the detector output is linear with respect to the LO power.
A homodyne receiver typically consists of a pair of high responsive linear photodiodes and linear transimpedance  amplifiers (TIA) \cite{Kumar2012}. The photo-diodes induce dark current noise which is, however, negligible  compare with the thermal noise  from  the  amplifier, the main electronic noise in the detector. 

In order to make the  homodyne detector  linear as well as shot noise limited it is required to use (i) a low-electronic noise amplifier with linear response and (ii) an adequate LO signal strength to  raise the  shot-noise variance well above the electronic noise of the amplifier. Note that the output variance of  the homodyne detection is   directly proportional to the  LO pulse  intensity that  interferes with the each of the signal pulses. In this sense, the number of photons per LO pulse is  responsible for the level of  shot-noise sensitivity  of the detector \cite{Kumar2012}.

The above mentioned two requirements set the  repetition rate and hence the secure key rate of the  CV-QKD system, in general. Most of the demonstrations of CV-QKD  have been limited to low repetition rate, of the order of MHz, as it is difficult to  provide low electronic noise and  high LO intensity  in high bandwidth homodyne receivers at high repetition rate. This implies that at high repetition rate the hardware assumptions to achieve theoretical security can not be met. Here we closely examin this aspect. The electronic noise variance, $v_{ele}$  of the homodyne receiver system is  governed  by the thermal noise  of the TIA. One can find  an expression for  $v_{ele}$ in shot noise unit as \cite{Laudenbach2018,Fuada2018}:

\begin{equation}
v_{ele} = \left(\frac{\sqrt{4kT_{k}R_f }}{G}\right)^2 \times \frac{BW d \lambda}{ h c P_{LO}}
 \label{JohnesonVele} 
\end{equation}

where, $k$ is the  Boltzmann constant, $T_k$ is the  temperature in kelvin, $R_f$ is the  feedback resistor  in the amplifier which is responsible for thermal noise, $G$ is the amplifier gain which decreases with increasing  bandwidth $BW$. The  term defined by the bracket  is referred to as the  noise equivalent power (NEP) at the input of the amplifier which can be directly obtained from the data sheet. In the rest of the Eq.\eqref{JohnesonVele}, \textit{h} is planck's constant, $c$ is the  speed of light and $\lambda, d, P_{LO}$ are wavelength,  pulse width and the power of the LO signal, respectively. In a high bandwidth CV-QKD realization, in order to keep the electronic noise significantly below the shot noise variance level, one has to use an amplifier with low NEP and maximum possible LO power though- both are very hard to be achieved in practice at high repetition rate.  This is because the NEP increases with bandwidth of the amplifier whereas the achievable LO power- more precisely  the number of photons per LO pulse, decrease with repetition rate as the pulse width decreases.

\begin{figure}[ht!]
\centering\includegraphics[width=0.8\textwidth]{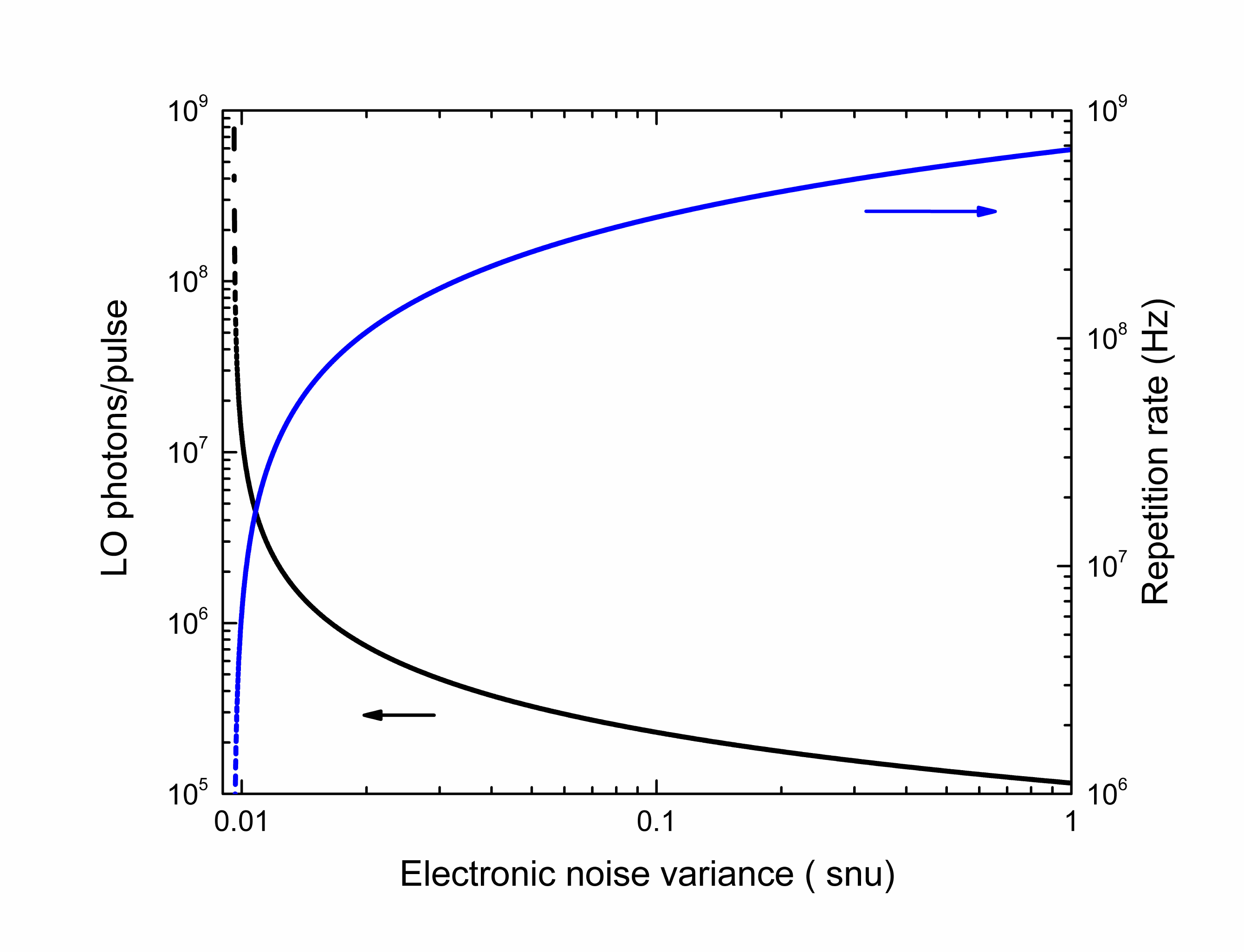}
\caption{Electronic noise variance, Eq.\eqref{JohnesonVele},with respect to repetition rate (blue) and  LO power (black). The following values are used: LO peak power - 1mW at the homodyne input, pulse width $d$ - $10 \%$ of the repetition rate, bandwidth BW- 3 times the repetition rate and $T_k$ - 300K. The gain G is set to 4000 at 1MHz and assumed linear decrease with increase in bandwidth.}
  \label{Vele} 
\end{figure}

As a result, elevated  electronic noise affects the  signal to noise ratio and effectively reduces the secure key rate. Fig.\ref{Vele} shows
variations in $v_{ele}$ with respect to repetition rate. In line with a common practice, here we set the  repetition rate is at one third of the amplifier bandwidth and the LO pulse width to be  $10\%$ duty cycle of the repetition rate.

As we can see from  Fig.\ref{Vele}, at high repetition rate $v_{ele}$ increases  due to decrease in LO power as well as increase in thermal noise. This results in the reduction in SNR and thereby the mutual information between Alice and Bob, as  per Eq.\eqref{Iab}.  This limits the achievable transmission distance and secure key rate.

In the following section we   describe muti-mode Gaussian modulated CV-QKD where a noisy homodyne detector can perform in the linear detection range with a reduction in electronic noise. This causes an elevation in shot noise sensitivity  with respect to the resultant virtual state - derived from the multi-mode signals, for secure key generation.

\section{Multi-mode Gaussian modulated CV-QKD}

\begin{figure}[htb!]
\centering\includegraphics[width=0.8\textwidth]{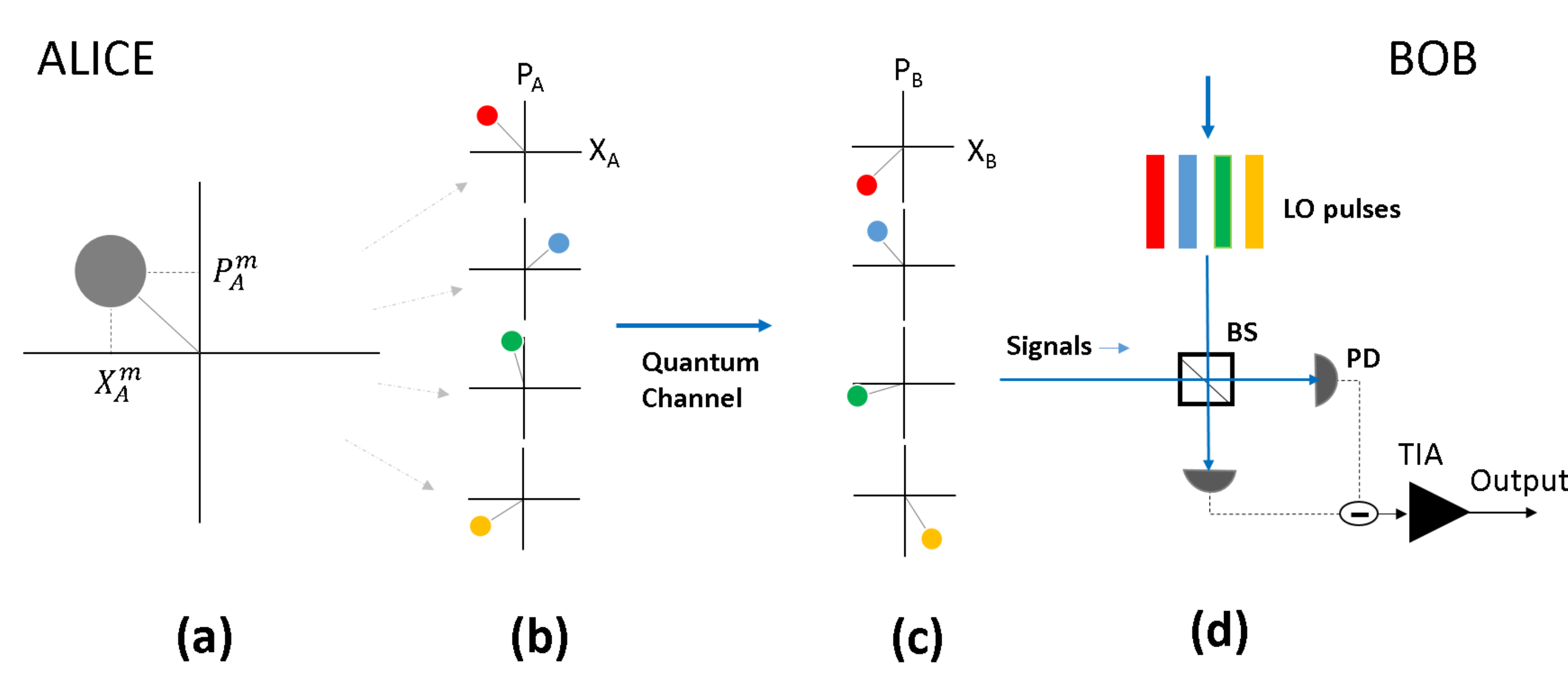}
 \caption{ Multi-mode CV-QKD Scheme. (a) the virtual coherent state of quadratures $X^m_A$ and $P^m_A$  deduced from the four signal mode quadratures. (b) 4 signal modes (coloured dots) are shown in respective phase space, with m=4 independent $X_A$ and $P_A$ quadrature values. Here, colour codes used for indicating  different modes.(c) Bob received the phase rotated signals. (d) Homodyne measurement of signal quadratures.  BS 50/50 beam splitter, PD photo-diode and TIA transimpedance amplifier. LO pulses pass through a phase modulator that is not shown in the figure.  }
\label{Setup}       
\end{figure}

In a conventional Gaussian modulated CV-QKD system Alice prepares a single mode coherent state $\ket{\alpha}$, where signal wavelength, polarization, spatial and temporal modes, etc., are coherent with respect to that of the LO pulse. In a multi-mode Gaussian modulated scheme, Alice prepares independent and identically distributed Gaussian modulated  coherent states  $\ket{\alpha}_1, \ket{\alpha}_2,. . ,\ket{\alpha}_m$ in $m$ independent modes as shown in Fig. \ref{Setup}(b). We assume modes are well separated from each other such that interference between the signals is negligible.

Bob then jointly measures, at once,  either of the quadratures of each mode using a single homodyne detector, with $m$ respective mode matched local oscillator pulses  as shown in Fig. \ref{Setup}(d). The joint outcome of  his $X$ quadrature measurement at any instance can be read as:

\begin{equation}
X^m_B = \sqrt{\eta T} \Sigma^m_{i=1} X_{A_i}
+ \Sigma^m_{i=1} X_{0_i}
+\sqrt{\eta T} \Sigma^m_{i=1} X_{\xi_i}
+ X_{ele}
 \label{MultimodeX} 
\end{equation}
Here,  $X_\xi, X_0$ and $X_{ele}$ are the  quadrature values of  excess noise, shot noise and electronic noise, respectively. Equation \eqref{MultimodeX} is also true for $P$ quadrature measurement.
Considering the additivity  of the Gaussian random variables, the variance $V_B^m = Var(X^m_B)$ of the measurement outcomes becomes:
\begin{equation}
V_B^m = \eta T m V_A +m N_0 + \eta T m\xi+ v_{ele}
 \label{MultimodeVAR} 
\end{equation} 
The excess noise variance  $\xi$ and  the  shot noise variance $N_0$ are assumed to be identical for all the modes. Since the shot noise variance at the homodyne output is proportional to the LO intensity, from a practical point of view, it  is not difficult to obtain LO pulses of identical strength and the same shot noise variance among the modes. While comparing Eq. \eqref{Var} of single mode  output variance with Eq. \eqref{MultimodeVAR}, we can see that the electronic noise $v_{ele}$ remains unchanged. This  reads as electronic noise per mode has been lowered by a factor of total number of modes, $m$. One may argue that the total noise, including the shot noise variance, is increased  by a factor m. However, normalization of  all the parameters with respect to total shot noise variance- in the resultant virtual phase space, the contribution of electronic noise is reduced by a factor m. 

In order to cope with Bob's  joint quadrature measurement, Alice estimates the resultant, but virtual, quadrature value $X^m_A = \Sigma^m_{i=1} X^i_A$ and $P^m_A = \Sigma^m_{i=1} P^i_A$  such that the resultant quadrature distribution still follows Gaussian distribution  $\mathcal{N}(0, m V_A)$ with mean zero and variance $mV_A$. A single virtual state is shown in Fig.\ref{Setup} (a). The uncertainty in the quadrature values are now considered as m times the shot noise variance $N_0$. However, this will be normalized to unity once joint shot noise variance measurement is done at Bob.

As  a result of the normalization with respect to the shot noise variance, $mN_0$,  the virtual states follow the Gaussian distribution $\mathcal{N}(0,  V_A^{m'})$ at Alice where $V_A^{m'}$ is identical to $V_A$. Here the superscript $m'$ labels the normalized virtual states. Exactly like in single mode protocol, Alice  publicly shares values of  her virtual quadratures and Bob estimates channel parameters, $T$ and $\xi$, which are common to all the modes, by following Eq.\eqref{Cov} and Eq.\eqref{MultimodeVAR}, 

\begin{eqnarray}
 T &=& \frac{\left<X_A^{m'} X_B^{m'}\right>^2}{\eta (V_A^{m'})^2} \\ \label{T} 
\xi &=& \frac{V_B^{m'}}{\eta T m} + V_A^{m'} + \frac{1}{\eta T} + \frac{v_{ele}}{\eta T m}
\end{eqnarray}

The SNR of the single mode CV-QKD system is defined as: $V_A/(1+\chi_{tot})$, where $\chi_{tot}$ is the  total noise in the system and its components are defined in  the  section 2. By taking them and Eq. \eqref{T}  into account, the ratio, $R_{snr}$, of SNR in multi-mode  to single mode Gaussian modulated CV-QKD system can be written as:  

\begin{equation}
R_{SNR} = \frac{1+\chi_{tot}}{1+\chi_{tot}/m + \xi(m-1)/m}
 \label{SNRs} 
\end{equation}

\begin{figure}[ht!]
\centering\includegraphics[width=\textwidth]{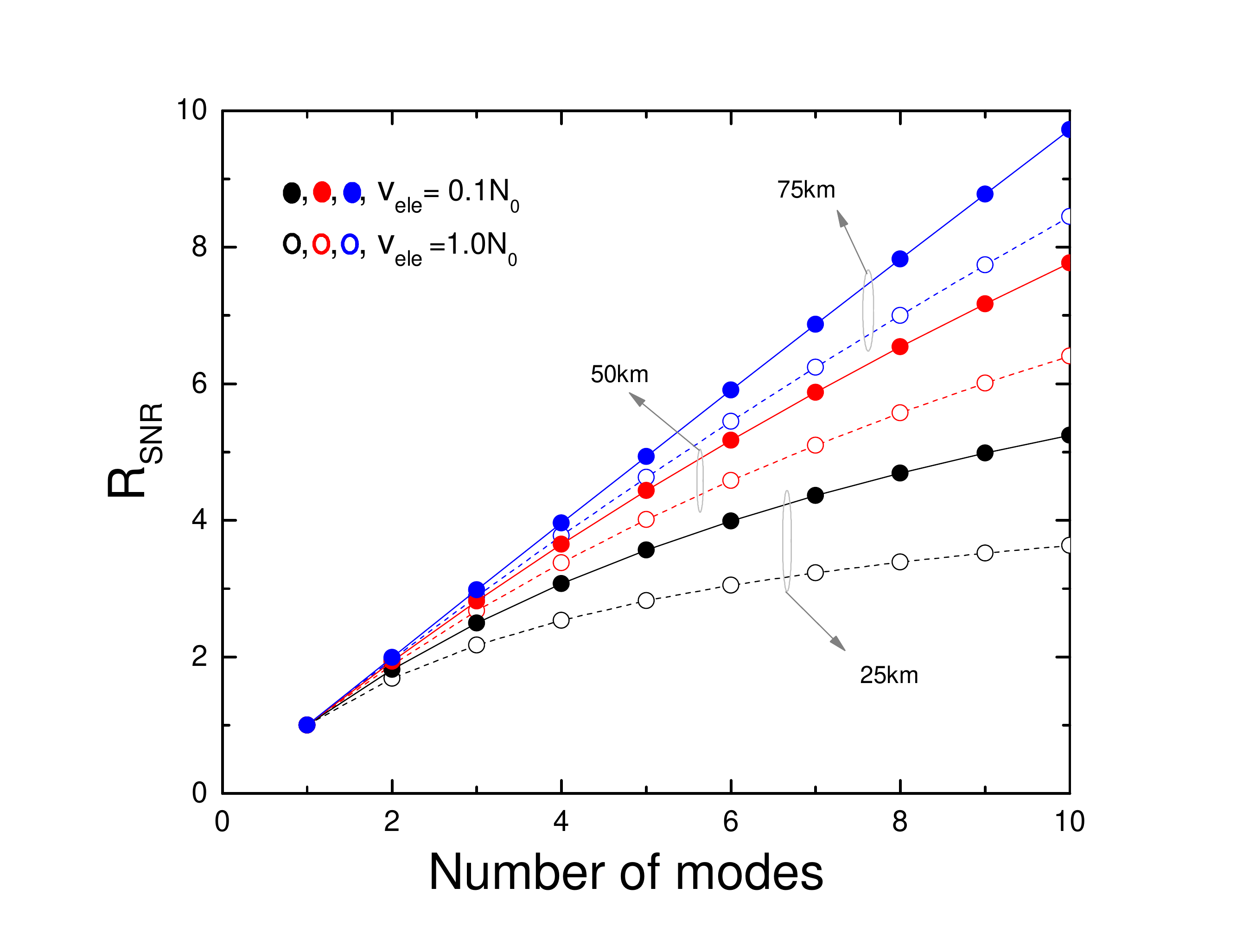}
\caption{The ratio $R_{SNR}$, Eq.\eqref{SNRs}, of multi-mode signal SNR to that of single mode. The following parameters are used  in the plot: single mode signal variance $V_A =2.5N_0$, reconciliation efficiency $\beta=0.95$, efficiency of Bob $\eta= 0.6$ and excess noise at Bob $\eta T \xi =0.001N_0$. Dots and circles represent the SNR ratio for $v_{ele}= 1.0N_0$ and $0.1N_0$, respectively.}
  \label{Rsnr} 
\end{figure}

The Eq.\eqref{SNRs} is plotted in Fig.\ref{Rsnr} in order to shows the enhancement in SNR of multi-mode  system.   
Here we assume that channel loss and detector  response to each of the modes are identical such that contribution of  each modes to the quadratures  of the virtual sates  are uniformly weighted. And also the quadrature variances of each modes can be set to identical during the system calibration procedure - which is a normal routine  preformed prior to the protocol run.  This lead us to the consider the  protocol with  final virtual states  follows  Gaussian linear model- where the  virtual states, the channel  noise, the  homodyne detection etc. are  identical to single mode Gaussian protocol. As a result, the security pertaining to the virtual state  can be derived  exactly as in the case of single mode Gaussian modulated coherent protocol \cite{Grosshans2005,Weedbrook2012a}. 

Therefore, in order to estimate the final secure key rate from the virtual states, the following equations are  utilized. The mutual information between Alice and Bob, $I_{AB}$, is estimated from Eq.\eqref{Iab} while  Eve's information  is calculated from  Eq.\eqref{CHIBE} given in appendix A. And finally utilizing Eq. \eqref{Keyrate}, we can estimate the secure key rate  generated from the virtual states. 
As usual, after the parameter  and key rate estimation Alice and Bob can proceed to error correction and then privacy amplification to generate final secure keys.

The LO pulses shown in Fig.\ref{Setup} (d) are either transmitted from Alice or locally generated at Bob. But, in LLO systems, it may not be necessary to  implement multi-mode scheme proposed here as it possible to generate adequate LO power from the local laser. Therefore we restrict our claims to the TLO systems. It is worth to mention here that in LLO system, noise from  phase estimation error misalign the reference frame that adds additional noise to the excess noise which  restricts the performance of  CV-QKD system to a few 10s of km\cite{Adrien2017}. For long distance CV-QKD, TLO scheme delivers secure keys as it is less vulnerable to phase estimation noise and our protocol finds application in high bandwidth and noisy detection. In order to align the frame of reference of the Alice's virtual quadrature values to  that of Bob's joint quadrature measurement outcomes, training signals can be  used for each modes with publicly known phase values\cite{Jouguet2011}. However, the  noise from reference frame misalignment  is not considered here.

In order to realize a full setup for  our protocol, the cost figure will be  primarily due to the lasers.  The modes can be generated with a set of lasers driven by single pulse generator, the laser outputs can be time delayed  and then coupled to a single fibre using a WDM module.  Gaussian quadrature modulation on each modes can be performed with single amplitude and phase modulators with  proper delay in the modulation pattern.  At Bob, an equal fibre delay can be applied to  each modes  in order to remove the delay offset by Alice.   The calibration procedure
  helps to  set equal weights on quadrature variance,  detector balancing , etc. of each modes.

\section{Proof of principle test and simulation results}
\begin{figure}[ht!]
\centering\includegraphics[width=0.8\textwidth]{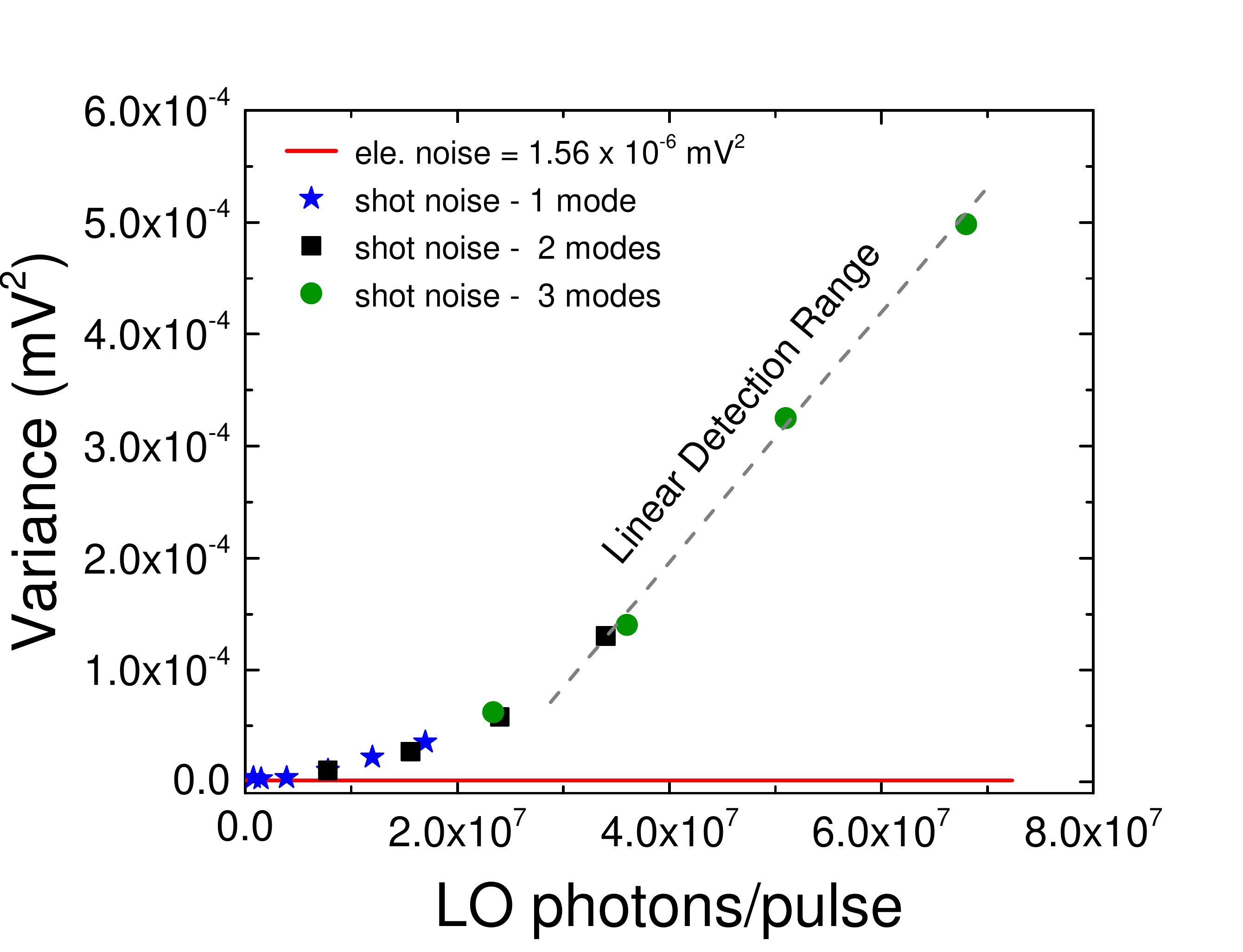}
\caption{Measured shot noise variance of the homodyne detector with LO power. Here, LO pulse width is 100ns at 1MHz repetition rate. With $1.7 \times 10^7$ photon per pulse, linear detection range can be reached with 3 LO pulses at which electronic noise per mode drops to $0.005N_0$. }
  \label{ShotNoise} 
\end{figure}

In order to demonstrate the  feasibility of a multi-mode GM-CVQKD system, we have used a 4MHz bandwidth homodyne detector that typically needs LO strength of $10^8$ photons per pulse to attain linear response to  the shot-noise quadrature and  thereby to the signal quadrature. Here, we  are restricted by the bandwidth of the available homodyne detector in the lab. However, it is possible to demonstrate the core of the concept with lower number of photon per  LO pulse where detector shows non-linear response  to the input signal and lesser shot noise sensitivity. This is exactly the scenario when higher bandwidth homodyne detectors show  non-linear response even at moderate LO power.
We test the  shot noise variance of the homodyne detector with 3 different modes of LO pulses, each of 100ns pulse width and wavelength 1544.53nm, 1545.32nm and 1546.12nm, respectively. Each LO generates its own contributions to the shot noise variance. The setup is shown as part of the Fig. \ref{Setup}.  The three CW  lasers are first multiplexed into a polarization maintaining fibre and then 100ns pulses are carved out with an amplitude modulator. The signal port of Bob's $50/50$ beamsplitter is blocked  and  LO pulses are sent simultaneously  though the LO input port.  The output currents from the photo-diodes are first subtracted and then amplified by low noise amplifier- Amptek A250. The output of the amplifier is acquired by a real time oscilloscope and processed in the computer, where variances of the homodyne output signals are estimated over $10^8$ data. Fig.\ref{ShotNoise} shows the  output variance of the detector for different LO pulses strength.

The electronic noise of the detector is $1.56 \times 10^{-6}$m$V^2$ which is measured  without the LO pulse.  The maximum  power for each LO mode is limited to $1.7 \times 10^7$  photons per pulse. While using single mode of LO pulse, the achievable minimum electronic noise is $\approx 0.05N_0$  whereas all the three LO pulses together bring the electronic noise variance down to $0.005N_0$- a 10 fold reduction, more than expected, which is due to non-linear gain of the amplifier. An immediate  implication of this demonstrations is that for  homodyne detectors which shows higher electronic noise, especially commercially available coherent receivers of GHz bandwidths, multi-mode detection can make them  sensitive to CV-QKD signals by significantly reducing the electronic noise variance with respect to shot noise variance.

To begin with, we  have estimated the secure key rate per  use of the channel  pertaining to  single mode Gaussian modulated CV-QKD protocol under collective attack, Eq.\eqref{Keyrate}  in two different homodye detector electronic noise levels $v_{ele}$: $0.1N_0$ and $1.0N_0$. One important thing to note here is that $v_{ele}=1N_0$  is a value commonly  observed for GHz bandwidth homodyne detectors, e.g 3GHz bandwidth detector at 1GHz repetition rate- see Fig.\ref{Vele}, and are not considered as shot noise limited. The value $0.1N_0$ is normally achievable in CV-QKD systems with repetition rate around 100MHz\cite{Duan2013} and detector can be regarded as  shot noise limited.

\begin{figure}[t]
\centering\includegraphics[width=0.8\textwidth]{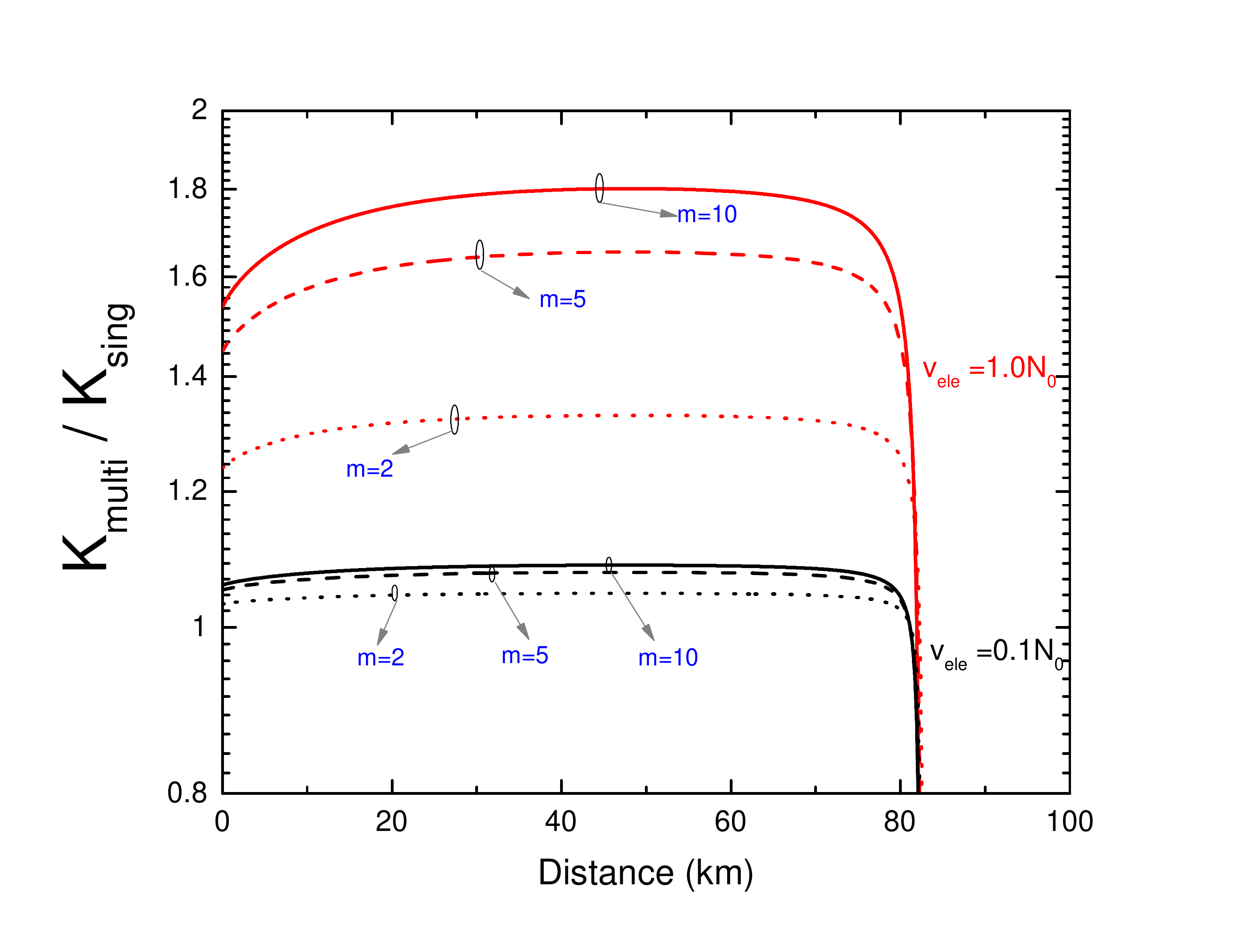}
\caption{Gain in terms of key rate in  multi-mode CV-QKD is shown for homodyne detector with low bandwidth (300MHz, $v_{ele}=0.1N_0$) and high bandwidth (3GHz, $v_{ele}=1N_0$). Here we used $V_A =2.5N_0$, reconciliation efficiency $\beta=0.95$, efficiency of Bob $\eta= 0.6$ and excess noise at Bob $\eta T \xi =0.001N_0$ and the channel loss = 0.2dB/km.  Red and  black colour in the plot indicate $v_{ele}$ = $1.0N_0$ and $0.1N_0$, respectively. }
  \label{Kratio} 
\end{figure}

A m-mode signal transmission can be considered as $m$ independent use of the channel which is equivalent to the transmission of $m$ single mode signal.  However, since Bob performs joint measurement on m-mode signals at once, the secure key of not equal to  m times the key rate from single mode transmission but higher. This  is shown in  Fig. \ref{Kratio} as $K_{multi}/K_{sing}$ which is the ratio  of the secure key rate  of multi-mode to single mode CV-QKD system for m = 2, 5 and 10. As we can see, muti-mode  scheme is less effective for CV-QKD system at lower repetition rate. But at higher bandwidth- above 1GHz, the multi-mode scheme brings the following advantages: i) improves the detector's  shot noise detection sensitivity by reducing the electronic noise variance per mode- here for m=10, $v_{ele} = 0.1N_0$; ii) it shows an increase in the secure key rate  with number of modes. For m=10, it almost doubles the key rate. One thing to mention here is that the numerical simulation does not consider  whether the detector is shot noise limited or not, but, estimates the key rate under the assumption that $v_{ele}$ is trusted and calibrated.

\section{Conclusion}

In conclusion, we have proposed a multi-mode Gaussian modulated CV-QKD scheme that conceptually reduces the electronic noise of the homodyne detector and provides shot noise sensitivity  during joint signal measurement.

 We have tested the principle behind the muli-mode system using a low bandwidth homodyne detector with low LO power and have observed a reduction in electronic noise, and an increase  in the shot noise sensitivity. This in turn increases the SNR of the CV-QKD system.  The ratio of secure key rates  of the conventional single mode Gaussian modulated CV-QKD and the multi-mode version has been estimated and found that  the multi-mode scheme has a negligible  impact  in low bandwidth CV-QKD systems while at higher bandwidth it provides shot noise sensitivity improvement and higher key rates compared to single mode systems. 
 
One of the hurdles for higher bandwidth Gaussian modulated CVQKD systems is the difficulty and latency in  data post-processing and hardware requirement for the   generation and acquisition of Gaussian random quadrature values. Discrete modulated CVQKD systems\cite{Leverrier2011}, with two, three or four signal levels,  reduce much of these complexities  and there are experimental demonstrations performed at GHz repetition rate \cite{Gunthner2017}.
The multi-mode scheme can not be applied to a discrete modulation scheme as the number of discrete signal levels of Alice's resultant virtual state as well as the result of Bob's joint measurement will increase with the number of modes. The security proofs developed for single mode discrete modulated protocols may not be directly applied here.  However, in the limit of large number of modes, the multi-mode discrete modulated scheme will converge to Gaussian modulation in the virtual quadrature under the central limit theorem.  This needs a careful study on whether security proofs for Gaussian modulation can apply to multi-mode discretely modulated CV-QKD protocols or not. We hope this paper will significantly add impact and give direction towards high bandwidth CV-QKD system development with currently unattainable repetition rates.

   \section*{Funding}Engineering and Physical Sciences Research Council (EPSRC) (EP/M013472/1). 

\section*{Appendix A: Eve's accessible information}
The security of prepare and measure based CV-QKD system has been  derived from the entanglement based scheme where Alice and Bob shares joint state $\rho_{AB} $. Alice's measurement on her state projects Bob's to a state $\alpha$ identical to what Alice would have sent to Bob in  prepare and measure scheme. The Eq.\eqref{CHIBE} is further simplified as:

\begin{equation}
\chi_{BE} = \sum_{i=1}^2 G\left(\frac{\lambda_i -1}{2} \right)- \sum_{i=3}^5 G\left(\frac{\lambda_i -1}{2} \right)
  \label{CHIBE1} 
\end{equation}

Where  $G(x) = (x+1)\log_2(x+1)-x\log_2 x$, $\lambda_{1,2}$ are the symplectic eigenvalues of the covariance matrix  that characterize a joint state $\rho_{AB}$ and $\lambda_{3,4,5}$ are that of the state left after Bob's measurement. One can find the eigenvalues as:

\begin{equation}
\lambda^2_{1,2} = \frac{1}{2}[A\pm\sqrt{A^2-4B}],
  \label{Lamba12} 
\end{equation}

in which, $A = V^2(1-2T)+2T+T^2(V+\chi_{line})^2$ and $B=T^2(V\chi_{line}+1)^2$ with $V=V_A+1$. Similarly,

\begin{equation}
\lambda^2_{3,4} = \frac{1}{2}[C\pm\sqrt{C^2-4D}],
  \label{Lamba34} 
\end{equation}
where, $C= ( V\sqrt{B}+T(V+\chi_{line})+A\chi_{ohm})/( T(V+\chi_{tot}))$ and $D= \sqrt{B}( (V+\sqrt{B}\chi_{ohm} )/( T(V+\chi_{tot} )))$  and the last symplectic eigenvalue  $\lambda_5$ is 1. Plugging Eq.\eqref{Lamba12} and Eq.\eqref{Lamba34} in Eq.\eqref{CHIBE1} we can estimate the upper bound of Eve's accessible information and then the final secure key rate from Eq.\ref{Keyrate}.

\bibliographystyle{plain}
\bibliography{cvref}
\end{document}